\documentclass[11pt,preprint]{aastex}

\usepackage{graphicx}

\shorttitle{Current-Driven Instability}
\shortauthors{Mizuno et al.}

\begin{document}

\title{3D Relativistic Magnetohydrodynamic Simulations of
Current-Driven Instability. I. Instability of a static column}

\author{
Yosuke Mizuno\altaffilmark{1,2}, Yuri Lyubarsky\altaffilmark{3}, Ken-Ichi
Nishikawa\altaffilmark{1,2}, and Philip E. Hardee\altaffilmark{4}}

\altaffiltext{1}{Center for Space
Plasma and Aeronomic Research, University of Alabama in Huntsville
,320 Sparkman Drive, NSSTC, Huntsville, AL 35805, USA; mizuno@cspar.uah.edu}
\altaffiltext{2}{National Space Science and Technology Center,
V62, Huntsville, AL 35805, USA}
\altaffiltext{3}{Physics Department, Ben-Gurion University, Beer-Sheva 84105, Israel}
\altaffiltext{4}{Department of
Physics and Astronomy, The University of Alabama, Tuscaloosa, AL
35487, USA}

\shorttitle{3D RMHD simulations of CD instability}
\shortauthors{Mizuno et al.}

\begin{abstract}

We have investigated the development of current-driven (CD) kink
instability through three-dimensional relativistic MHD simulations. A
static force-free equilibrium helical magnetic configuration is
considered in order to study the influence of the initial
configuration on the linear and nonlinear evolution of the
instability. We found that the initial configuration is strongly
distorted but not disrupted by the kink instability.  The instability
develops as predicted by linear theory.  In the non-linear regime the
kink amplitude continues to increase up to the terminal simulation
time, albeit at different rates, for all but one simulation.  The
growth rate and nonlinear evolution of the CD kink instability depends
moderately on the density profile and strongly on the magnetic pitch
profile.  The growth rate of the kink mode is reduced in the linear
regime by an increase in the magnetic pitch with radius and the
non-linear regime is reached at a later time than for constant helical
pitch. On the other hand, the growth rate of the kink mode is
increased in the linear regime by a decrease in the magnetic pitch
with radius and reaches the non-linear regime sooner than the case
with constant magnetic pitch. Kink amplitude growth in the non-linear
regime for decreasing magnetic pitch leads to a slender helically
twisted column wrapped by magnetic field. On the other hand, kink
amplitude growth in the non-linear regime nearly ceases for increasing
magnetic pitch.

\end{abstract}

\section{Introduction}

Relativistic jets occur in active galactic nuclei ({\bf AGN}) (e.g.,
Urry \& Padovani 1995; Ferrari 1998; Meier et al. 2001), occur in
microquasars (e.g., Mirabel \& Rodriguez 1999), and are thought
responsible for the gamma-ray bursts ({\bf GRB}) (e.g., Zhang \&
M\'{e}sz\'{a}ros 2004; Piran 2005; M\'{e}sz\'{a}ros 2006).  The most
promising mechanism for producing relativistic jets involves
magnetohydrodynamic acceleration from an accretion disk around a black
hole (e.g., Blandford 1976; Lovelace 1976; Blandford \& Payne 1982;
Fukue 1990; Meier 2005; Narayan et al. 2007), and/or involves the
extraction of energy from a rotating black hole (Penrose 1969;
Blandford \& Znajek 1977).

General relativistic magnetohydrodynamic ({\bf GRMHD}) codes have been
used to study the extraction of rotational energy from a spinning
black hole, i.e., Blandford-Znajek mechanism (Koide 2003; Komissarov
2005; McKinney 2005; Komissarov \& McKinney 2007; Komissarov \& Barkov
2009) and from an accretion disk, i.e., Blandford-Payne mechanism
(Blandford \& Payne 1982). GRMHD codes also probe the formation of GRB
jets in collapsars (e.g., Mizuno et al.\ 2004a, 2004b; Liu et
al. 2007; Barkov \& Komissarov 2008; Stephens et al. 2008; Nagataki
2009; Komissarov \& Barkov 2009) and are used to study the propagation
of jets injected from an unresolved Keplerian disk (e.g., Komissarov
et al. 2007, 2008; Tchekhovskoy et al. 2008, 2009).  GRMHD simulations
with a spinning black hole indicate jet production with a magnetically
dominated high Lorentz factor spine with $v \sim c$, and a matter
dominated sheath with $v \gtrsim c/2$ possibly embedded in a lower
speed, $v \ll c$, disk/coronal wind (e.g., De Villiers et al. 2003,
2005; Hawley \& Krolik 2006; McKinney \& Gammie 2004; McKinney 2006;
McKinney \& Narayan 2007a, 2007b; McKinney \& Brandford 2008; Beckwith
et al.\ 2008; Hardee et al.\ 2007).  Circumstantial evidence such as
the requirement for large Lorentz factors suggested by the TeV BL Lacs
when contrasted with much slower observed motions (Ghisellini et al.\
2005) suggests such a spine-sheath morphology, although alternative
interpretations are also possible (Georganopoulos \& Kazanas 2003;
Bromberg \& Levinson 2008; Stern \& Poutanen 2008; Giannios et al.\ 2009). 

Numerical simulations indicate that the jet spine and sheath are
accelerated and, in part, collimated by strong magnetic fields twisted
in the rotating black hole ergosphere and in the accretion disk,
respectively. The helically twisted magnetic fields are expected to be
dominated by the toroidal component in the far zone as a result of jet
expansion because the poloidal component falls off faster with
expansion and distance. In configurations with strong toroidal
magnetic field the current driven ({\bf CD}) kink mode is
unstable. This instability excites large-scale helical motions that
can strongly distort or even disrupt the system.  For static
cylindrical force-free equilibria, the well-known Kruskal-Shafranov
criterion states that the instability develops if the length of the
column, $\ell$, is long enough for the field lines to go around the
cylinder at least once (e.g. Bateman 1978): $|B_{p}/B_{\phi}| < \ell/ 2
\pi R$. This criterion suggests that jets are unstable beyond the 
Alfv\'{e}n surface
because of $B_{p} \la B_{\phi}$ and $\ell \ga R_{j}$ at the Alfv\'{e}n
surface, where $R_{j}$ is the cylindrical radius of the jet.  However,
rotation and shear motions could significantly affect the instability
criterion.  For relativistic force-free configurations, the linear
instability criteria were studied by Istomin \& Pariev (1994, 1996),
Begelman (1998), Lyubarskii (1999), Tomimatsu et al.\ (2001) and
Narayan et al.\ (2009).

The linear mode analysis provides conditions for the instability but
says little about the impact the instability has on the system. The
instability of the potentially disruptive kink mode found from a
linear analysis must be followed into the non-linear regime.  The
nonlinear development of the CD kink instability is different for
external and internal kink modes (e.g., Bateman 1978). External
disruptive instability develops in a current carrying plasma column
surrounded by a vacuum magnetic field. In astrophysical systems, some
amount of plasma is present everywhere. In this case an internal kink
instability develops inside a resonant surface on which the helicity
of the magnetic field lines matches that of the eigenmode. 
In dissipationless MHD, the internal instability is known to saturate.
Finite resistivity makes the internal kink mode disruptive but the time-scale
significantly exceeds the Alfv\'{e}n time-scale and may be too long to
disrupt a fast outflow. In this case helical structures may be formed
in the flow.  Recently helical structures have been found in
non-relativistic/relativistic simulations of magnetized jets (e.g.,
Lery et al.\ 2000; Ouyed et al.\ 2003; Nakamura \& Meier 2004;
Nakamura et al.\ 2007; Moll et al.\ 2008; McKinney \& Blandford 2008;
Carey \& Sovinec 2009).

Twisted structures are observed in many AGN jets on sub-parsec, parsec
and kiloparsec scales (e.g., G\'{o}mez et al.\ 2001; Lobanov \& Zensus
2001). Dissipation processes may result in relaxation to helical
equilibrium as evoked by K\"{o}nigl \& Choudhuri (1985) for helical
structures in AGN jets.  In the absence of CD kink instability and
resistive relaxation, helical structures may be attributed to the
Kelvin-Helmholtz ({\bf KH}) helical instability driven by velocity
shear at the boundary between the jet and the surrounding medium
(e.g., Hardee 2004, 2007) or triggered by precession of the jet
ejection axis (Begelman et al.\ 1980).  It is still not clear whether
current driven, velocity shear driven or jet precession is responsible
for the observed structures, or whether these different processes are
responsible for the observed twisted structures at different spatial
scales.

This is the first in a series of papers in which we study kink
instability in relativistic systems. By relativistic we mean not only
relativistically moving systems but any with magnetic energy density
comparable to or greater than the plasma energy density, including the
rest mass energy. In this paper, we present 3D results of the CD kink
instability of a static plasma column. We start from static
configurations because in the case of interest, the source of the free
energy is the magnetic field, not the kinetic energy. Therefore static
configurations (or more generally rigidly moving flows considered in
the proper reference frame) are the simplest ones for studying the
basic properties of the kink instability. At the next stage, we will
investigate the influence of shear motions and rotation on the
stability and nonlinear behavior of jets. This article is organized as
follows.  We describe the numerical method and setup used for our
simulations in \S 2, present our results in \S 3, discuss the
astrophysical implications in \S 4, and present some numerical tests
in the Appendix.

\section{Numerical Method}

In order to study time evolution of the CD kink instability in the
relativistic MHD ({\bf RMHD}) regime, we use the 3D GRMHD code
``RAISHIN'' in Cartesian coordinates. RAISHIN is based on a $3+1$
formalism of the general relativistic conservation laws of particle
number and energy momentum, Maxwell's equations, and Ohm's law with no
electrical resistance (ideal MHD condition) in a curved spacetime
(Mizuno et al.\ 2006).  In the RAISHIN code, a conservative,
high-resolution shock-capturing scheme is employed. The numerical
fluxes are calculated using the HLL approximate Riemann solver, and
flux-interpolated, constrained transport is used to maintain a
divergence-free magnetic field. The RAISHIN code has proven to be
accurate to second order and has passed a number of numerical tests
including highly relativistic cases and highly magnetized cases in
both special and general relativity (Mizuno et al.\ 2006). The RAISHIN
code performs special relativistic calculations in Minkowski spacetime
by changing the metric.

For our simulations we will choose a force-free helical magnetic field
as the initial configuration. A force-free configuration is a
reasonable choice for the strong magnetic field cases that we study
here.  In general, the force-free equilibrium of a static cylinder is
described by the equation
\begin{equation}
B_{z} {d B_{z} \over dR} + {B_{\phi} \over R} {d B_{\phi} \over dR} =0.
\end{equation}
In particular we choose a poloidal magnetic field component of the form
\begin{equation}
B_{z}= {B_{0} \over [1+ (R/a)^{2}]^{\alpha}}~,
\end{equation}
for which one finds a toroidal magnetic field component of the form
\begin{equation}
B_{\phi}= {B_{0} \over (R/a)[1+ (R/a)^{2}]^{\alpha}} \sqrt{ { [1 +
(R/a)^{2}]^{2 \alpha} -1 - 2 \alpha (R/a)^2 \over 2 \alpha -1}}~,
\end{equation}
where $R$ is the cylindrical radial position normalized by a
simulation scale unit $L \equiv 1$, $B_{0}$ parameterizes the magnetic
field amplitude, $a$ is the characteristic radius of the column, and
$\alpha$ is the pitch profile parameter.  The pitch profile parameter
determines the radial profile of the magnetic pitch $P=R B_{z} /
B_{\phi}$, and provides a measure of the twist of the magnetic field
lines. With our choice for the force-free field, the magnetic pitch
can be written as
\begin{equation}
P = (R/a)^{2} \sqrt{ { 2 \alpha -1 \over [1 + (R/a)^{2} ]^{2 \alpha}
-1 -2 \alpha (R/a)^{2}} }~.
\end{equation}
If the pitch profile parameter $\alpha < 1$, the magnetic pitch
increases with radius.  If $\alpha > 1$, the magnetic pitch
decreases. When $\alpha =1$, the magnetic pitch is constant.  This
configuration is the same as that used in previous non-relativistic
work (Appl et al.\ 2000; Baty 2005).

The simulation grid is periodic along the axial direction. As a
consequence the allowed axial wavelengths are restricted to $\lambda =
L_{z}/n \le L_{z}$, with $n$ a positive integer and $L_{z}$ is the
grid length. The grid is a Cartesian ($x, y, z$) box of size $4L
\times 4L \times L_{z}$ with grid resolution of $\Delta L = L/40$. In
simulations we choose two different column radii, $a$, relative to the
column length, $L_{z}$: (case A) $a=(1/16)L_{z}=(1/8)L$ and (case B)
$a=(1/12)L_{z}=(1/4)L$. For case A $L_{z} = 2L$ and for case B $L_{z} = 3L$.
In terms of $a$, the simulation box size is
$32a \times 32a \times 16a$ for case A and $16a \times 16a \times 12a$
for case B.  We impose outflow boundary conditions on the transverse
boundaries at $x = y = \pm 2L$ ($x = y = \pm 16a$ for case A and $x =
y = \pm 8a$ for case B).

We consider a low gas pressure medium with constant $p=p_{0}
=0.02\rho_{0}c^{2}$ for the equilibrium state, and two different
density profiles: (case u) uniform density $\rho = 1.0 \rho_{0}$ and
(case n) non-uniform density decreasing proportional to the magnetic field
strength, $\rho = \rho_{1} B^{2}$ with $\rho_{1} =10.0\rho_{0}$. The
equation of state is that of an ideal gas with $p=(\Gamma -1) \rho e$,
where $e$ is the specific internal energy density and the adiabatic
index $\Gamma=5/3$. The specific enthalpy is $h \equiv 1+e/c^{2}
+p/\rho c^{2}$. The magnetic field amplitude is $B_{0}
=0.4\sqrt{4\pi\rho_{0}c^{2}}$ leading to a low plasma-$\beta$ near the
axis. The sound speed is $c_{s}=(\Gamma p/\rho h)^{1/2}$
and the Alfv\'{e}n speed is $v_{A} =[B^{2}/(\rho h +B^{2})]^{1/2}$.

In order to investigate different radial pitch profiles, we perform
simulations with: (case 1) constant pitch, $\alpha=1$, (case 2)
increasing pitch, $\alpha=0.75$, and (case 3) decreasing pitch
$\alpha=2.0$. The radial profiles of the magnetic field components,
the magnetic pitch, and the sound and Alfv\'en speeds for the different
cases are shown in Figure 1. 
The radial profile of the magnetic field components for cases A and B
are the same when normalized by $a$.
\begin{figure}[h!]
\epsscale{1.0}
\plotone{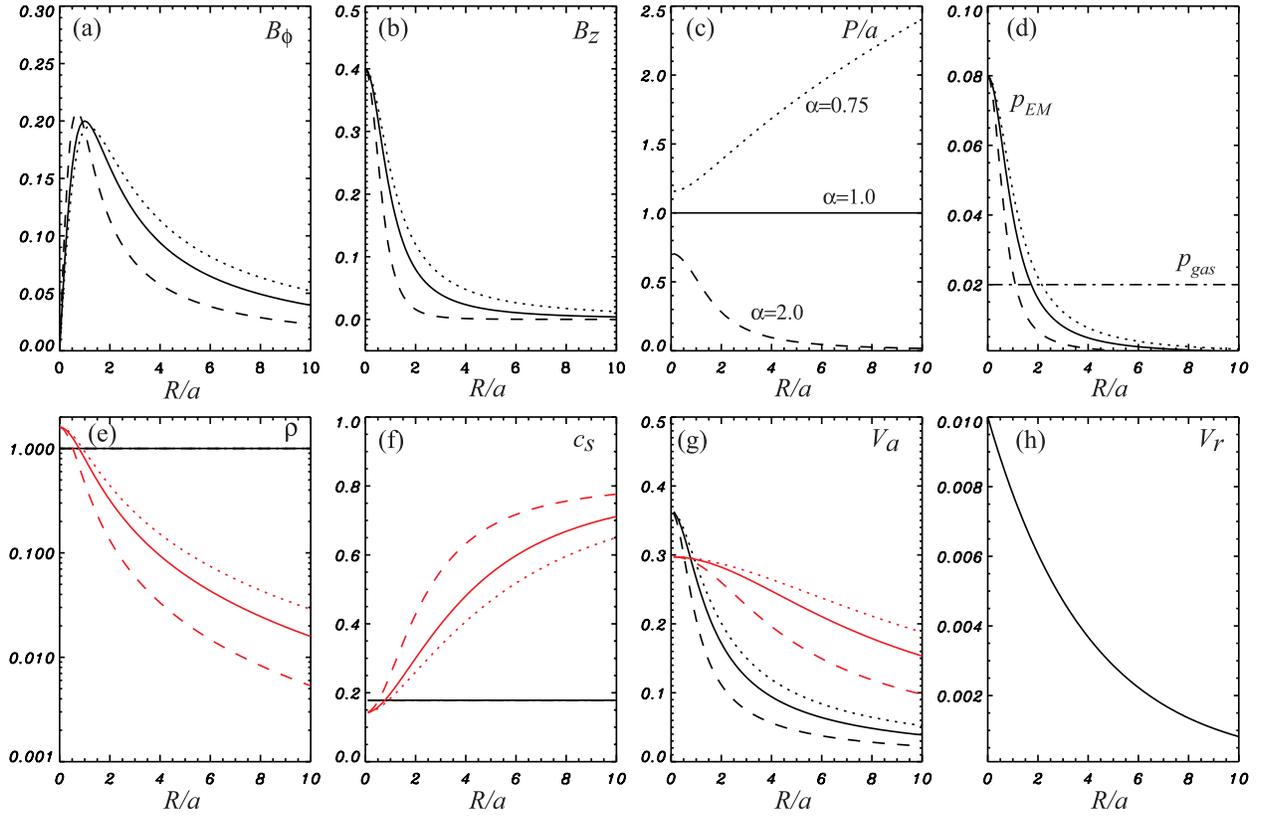}
\caption{Radial profile of (a) the toroidal magnetic field
($B_{\phi}$),(b) the axial magnetic field ($B_{z}$), (c) the magnetic
pitch, $P=R B_{z}/B_{\phi}$, (d) magnetic and gas (dash-dot line)
pressure, (e) the rest mass density, (f) the sound speed, and (g) the
Alfv\'{e}n speed. Uniform density cases are in black and declining
density cases are in red where (solid) indicates constant pitch,
(dotted) indicates increasing pitch, and (dashed) indicates decreasing
pitch. The radial velocity perturbation profile is shown in panel (h).
\label{f1}}
\end{figure}

The initial MHD equilibrium configuration is perturbed by a small
radial velocity with profile given by
\begin{equation}
v_{R} = \delta v \exp \left( - { R \over R_{a} } \right) \cos (m
\theta) \sin \left( {2 \pi n z \over L_{z}}\right)~.
\end{equation}
The amplitude of the perturbation is taken to be $\delta v = 0.01c$
with radial width $R_{a}= 0.5L$ ($4a$ for case A and $2a$ for case
B). We choose $m=1$ and $n=1$ in the above formula.  This is identical
to imposing $(m,n)=(- 1, -1)$, because of the symmetry between $(m,n)$
and $(-m, -n)$ pairs. The various different cases that we have
considered are listed in Table 1.
\begin{deluxetable}{lccccccc}
\tablecolumns{7}
\tablewidth{0pc}
\tablecaption{Models and Parameters}
\label{table1}
\tablehead{
\colhead{Case} & \colhead{$a/L$} & \colhead{$\alpha$} &
\colhead{$\rho$} & \colhead{Pitch} & \colhead{$L_{z}/L$} & 
\colhead{$L_{z}/a$} & \colhead{$v_{A0}/c$}
} 
\startdata
A1u &  0.125  &  1.0  &  uniform  &  constant  &  2.0 & 16 & 0.4 \\
A1n &  0.125  &  1.0  &  decrease & constant & 2.0 & 16 & 0.3 \\
B1u &  0.25   &  1.0  &  uniform & constant & 3.0 & 12 & 0.4 \\
B1n &  0.25   &  1.0  &  decrease & constant & 3.0 & 12 & 0.3 \\
A2u &  0.125  &  0.75 & uniform & increase & 2.0 & 16 & 0.4 \\
A3u &  0.125  &  2.0  & uniform & decrease & 2.0 & 16 & 0.4 \\
B2u &  0.25   &  0.75 & uniform & increase & 3.0 & 12 & 0.4 \\
B3u &  0.25   &  2.0  & uniform & decrease & 3.0 & 12 & 0.4 \\
A2n &  0.125  &  0.75 & decrease & increase & 2.0 & 16 & 0.3 \\
A3n &  0.125  &  2.0  & decrease & decrease & 2.0 & 16 & 0.3 \\
B2n &  0.25   &  0.75 & decrease & increase & 3.0 & 12 & 0.3 \\
B3n &  0.25   &  2.0  & decrease & decrease & 3.0 & 12 & 0.3 \\

\enddata
\end{deluxetable}

\clearpage

\section{Results}

\subsection{Time evolution of the CD kink instability}

As an indicator of the growth of the CD kink instability we use
the volume-averaged kinetic energy transverse to the $z$-axis
within a cylinder of radius $R/L \le 1.0$ written as
\begin{equation} E_{kin,xy}={1
\over V_{b}} \int_{V_b} {\rho v^{2}_{x} + \rho v^{2}_{y} \over 2} dx
dy dz
\end{equation}
and normalized by the initial volume-averaged transverse magnetic
energy. The quantity
$E_{kin,xy}$ allows determination of different evolutionary
stages (initial exponential (linear growth phase) growth, and subsequent
non-linear evolution). In all cases, the initial growth
regime is characterized by an exponential increase in $E_{kin,xy}$ by
several orders of magnitude to a maximum amplitude followed
by a slow decline in the non-linear regime. The time evolution of the
maximum radial velocity evaluated over half the grid length,
$L_{z}/2$, shows a growth trend similar to the growth of $E_{kin,xy}$.

\begin{figure}[h!]
\epsscale{0.85}
\plotone{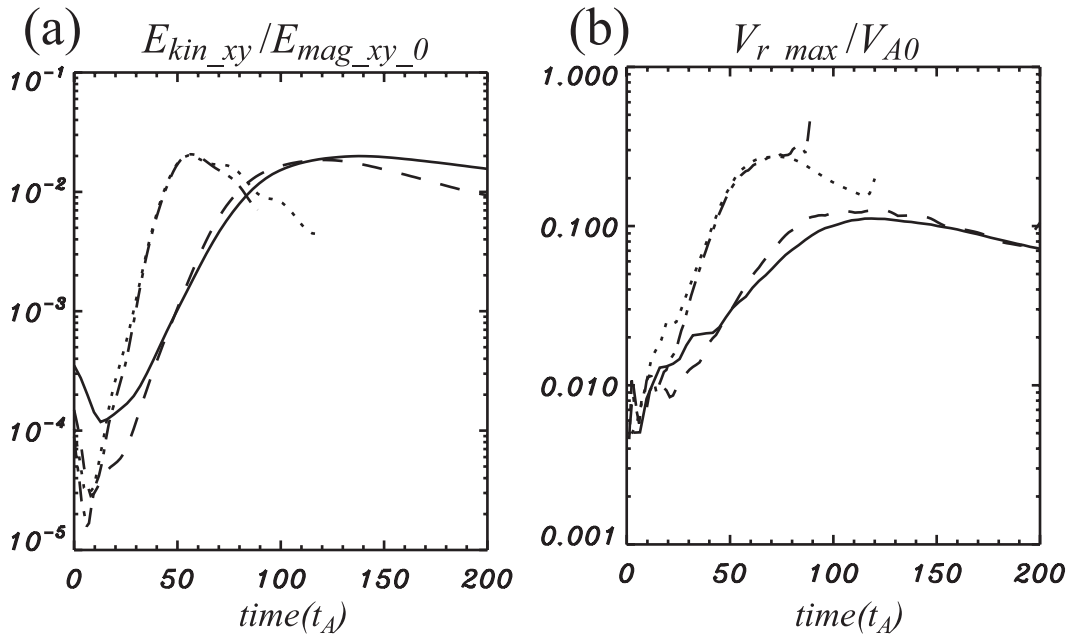}
\caption{Time evolution of (a) $E_{kin,xy}$ within $R/L \le 1.0$ normalized by the initial volume-averaged magnetic energy and (b) the
maximum radial velocity normalzed by the Alfv\'{e}n velocity on the
axis ($v_{A0}$) for constant pitch ($\alpha=1.0$). Case A,
$a=(1/16)L_{z}$: uniform density (solid line) and decreasing density
(dotted line). Case B, $a=(1/12)L_{z}$: uniform density (dashed line)
and decreasing density (dash-dotted line). Time is in units of $t_A =
a/v_{A0}$, where $v_{A0}$ is initial Alfv\'{e}n velocity on the
axis. Note that normalizing values of $E_{mag,xy,0}$, $v_{A0}$ and the
timescale units, $t_A$, are different for uniform and decreasing
density cases.\label{f2}}
\end{figure}
Figure 2 shows the time evolution of $E_{kin,xy}$ and maximum radial
velocity for cases A and B with constant pitch for uniform and
decreasing radial density profiles.  The effect of changing the
characteristic radius, $a$, relative to the grid length, $L_{z}$, is
equivalent to changing the wavelength. The wavelength in case A is
$\lambda = 16~a$, and in case B is $\lambda = 12~a$. Note that
according to the Kruskal-Shafranov criterion, the instability
developes at $\lambda>2\pi a$. The instability growth rate reaches a
maximum at $\lambda_{max}\approx 10~a$, the exact coefficient being
dependent on the transverse distribution of the density and magnetic
pitch. Specifically in the case of constant pitch and uniform
density, Appl et al.\ (2000) found $\lambda_{max} = 8.43~a$ and a
corresponding growth rate of $\Gamma_{max}=0.133~v_{a0}/a$.  In general,
one can use the estimate $\Gamma_{max}\approx 0.1~v_{a0}/a$.

In uniform density (solid and dashed lines in Fig.\ 2) cases
$E_{kin,xy}$ grows exponentially (linear growth phase), reaches
maximum amplitude at $t_{S} \sim 100 t_{Au}$ for $a=(1/16)L_{z}$ and
at $t_{S}\sim 90 t_{Au}$ for $a=(1/12)L_{z}$, and slowly declines in
the non-linear regime.  The timescale $t$ is in units of the
Alfv\'{e}n crossing time of the characteristic radius $a$, i.e.,
$t_{Au} \equiv a/v_{A0}$ where the initial Alfv\'{e}n velocity on the
axis is $v_{A0}\simeq 0.4~c$. The difference in growth is a result of
the different wavelengths, longer in case A and shorter in case B.
Both cases show similar amplitude of $E_{kin,xy}$ at transition to the
non-linear stage.  Here we find that the shorter wavelength, $\lambda
= 12~a$, grows slightly more rapidly than the longer wavelength,
$\lambda = 16~a$, as would be expected if $\lambda_{max} < 12~a$.

In decreasing density cases (dotted lines and dash-dotted lines in
Fig.\ 2), the initial growth in the linear growth phase is more rapid
and reaches maximum amplitude at $t_{S} \sim 50 t_{An}$ for
$a=(1/16)L_{z}$ and for $a=(1/12)L_{z}$, and slowly declines in the
non-linear regime.  Here the initial Alfv\'{e}n velocity on the axis
is $v_{A0} \simeq 0.3~c$ and it follows that $t_{An} = (4/3) t_{Au}$.
This more rapid growth is a result of dependence of the growth rate on
the Alfv\'{e}n velocity. The density decline in the decreasing case
leads to a more gradual radial decline in the Alfv\'{e}n velocity (see
Fig.\ 1g) than for the uniform density cases, and on average a higher
Alfv\'en velocity. The maximum amplitude of $E_{kin,xy}$ relative to
the initial volume-averaged magnetic energy for the decreasing density
cases is almost the same as the uniform density cases. The maximum
radial velocities normalized by the initial axial Alfv\'en velocity
are higher for the decreasing density cases but in absolute terms are
almost the same. We note that the normalized maximum radial velocity
in the decreasing density cases appears larger than in the uniform
density cases because the normalizing axial Alfv\'{e}n velocity is
smaller for the decreasing density cases.

\begin{figure}[hp]
\epsscale{0.85}
\plotone{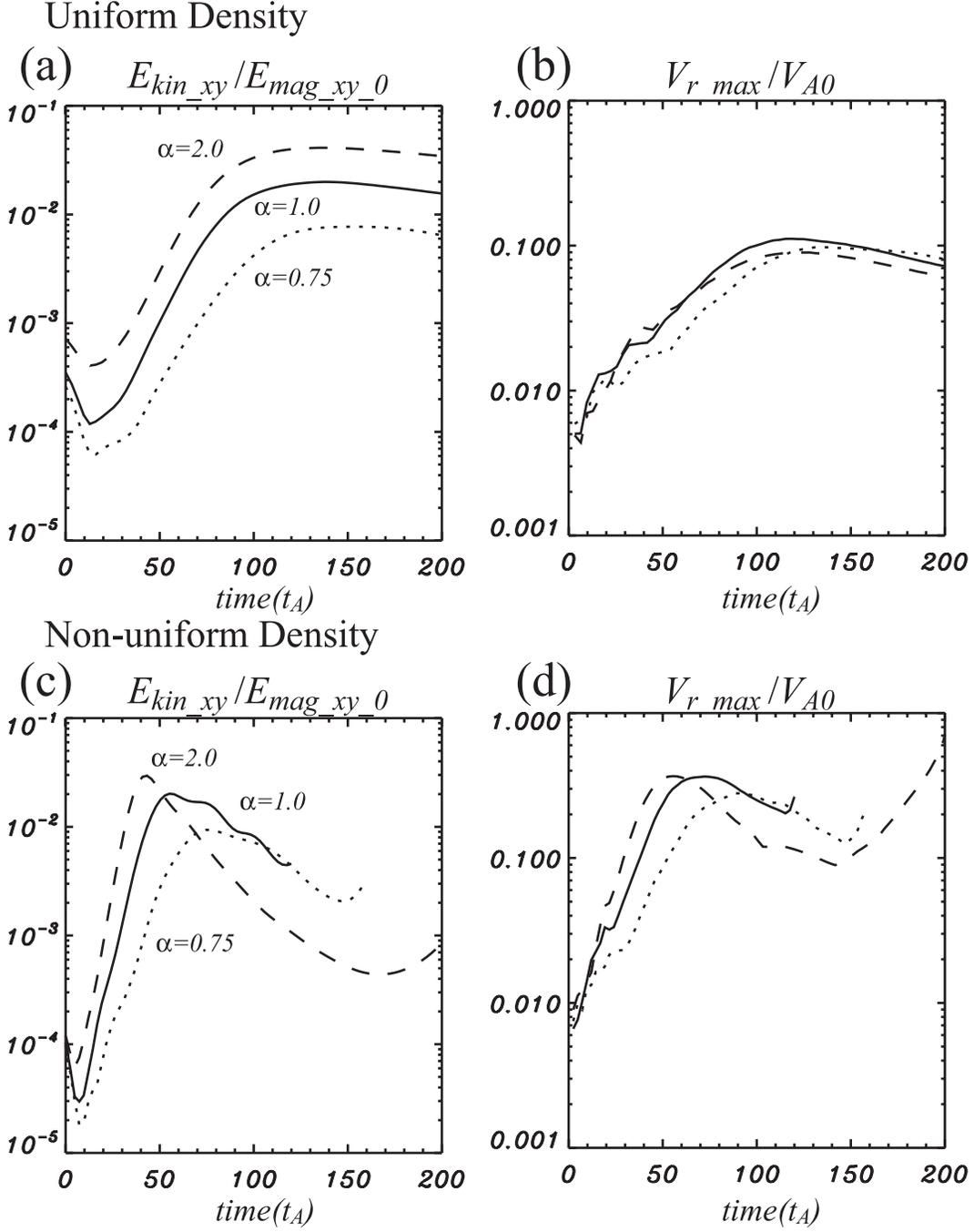}
\caption{Time evolution of (a,c) $E_{kin,xy}$
within $R/L \le 1.0$ normalized by the initial volume-averaged
magnetic energy and (b,d) maximum radial velocity normalzed by the
initial Alfv\'{e}n velocity on the axis ($v_{A0}$). Upper panels show
the uniform and lower panels show the decreasing (lower) density cases
with constant pitch $\alpha=1.0$ (solid lines), increasing pitch
$\alpha=0.75$ (dotted lines), and decreasing pitch $\alpha=2.0$
(dashed lines) cases for $a=(1/8) L$. Time is in units of $a/v_{A0}$,
where $v_{A0}$ is initial Alfv\'{e}n velocity on the axis. Note that
normalizing values of $E_{mag,xy,0}$, $V_{A0}$ and the timescale
units, $t_A$, are different for uniform and decreasing density
cases. \label{f3}}
\end{figure}
The effect of different radial pitch on the time evolution of $E_{kin,
xy}$ and the maximum radial velocity for cases with different radial
pitch are shown in Figure 3.  The three uniform density and three
decreasing density cases show similar linear and non-linear evolution
but with different time scales.  Growth in the linear stage is more
rapid for the decreasing density cases than for the uniform density
cases. This can be attributed to the higher average Alfv\'en
velocity in the decreasing density case.  For both uniform and
decreasing density cases the increasing pitch case ($\alpha = 0.75$)
grows more slowly and reaches maximum at a later time with a smaller
value for $E_{kin,xy}$ than the constant pitch case ($\alpha =
1.0$). On the other hand, the decreasing pitch case ($\alpha = 2.0$)
grows more rapidly and reaches maximum at an earlier time with a
larger value for $E_{kin,xy}$ than the constant pitch case. Although
the transition time from linear to non-linear evolution is different
for each pitch case, the maximum radial velocity is almost the same at
transition. The different growth rates as a function of the radial
pitch profile are consistent with the non-relativistic linear analysis
in Appl et al.\ (2000).

\newpage
\vspace{-0.5cm}
\subsection{Three dimensional structure of the CD kink instability}

The results of the previous subsection show that the instability
excites relatively slow motions, $v \ll v_A$, and that the overall
energy of the system does not vary too much. Nevertheless, the
three-dimensional structure of the system is strongly distorted, as
will be shown in this subsection. We note that results for case B with
the shorter wavelength are essentially the same as for case
A. Moreover, the evolution is similar in the uniform and decreasing
density cases and only occurs faster in the decreasing density
cases. Therefore we show the results of case A with decreasing density
only.

Figure 4 shows the time evolution of a density isosurface for the
constant pitch case A1n.  Displacement of the initial force-free
helical magnetic field leads to a helically twisted magnetic filament
around the density isosurface.  At longer times, the radial
displacement of the high density region, best seen in transverse
density slices at the grid midplane $z = L_{z}/2$ (see right panels in
Fig.\ 4), slows significantly. Continuing outwards radial motion is
confined to a lower density sheath around the high density core.

The transverse growth of helical twisting is illustrated by the time
evolution of magnetic field line displacement seen from the pole-on
view shown in Figure 5. Here we follow field lines anchored initially
at $R/L=0.2$ at the $z=0$ surface. Displacement of the magnetic field
rapidly increases with time in the linear stage and the displacement
continues to gradually increase throughout the non-linear phase.  Thus,
the magnetic field displacement indicates continued growth even though
the high density region has ceased significant outwards motion by the
terminal simulation time.  Evolution of the helical perturbation in
the magnetic field is accompanied by an appropriate evolution of the
current distribution. Initially a large axial current is located near
the axis. In the linear growth phase, the axial current is displaced
and twists helically.  The axial current twist is cospatial with the
high density twist.  In the linear growth phase the axial current
gradually decreases and in the non-linear phase remains constant in
magnitude.

\begin{figure}[hp]
\epsscale{0.8}
\plotone{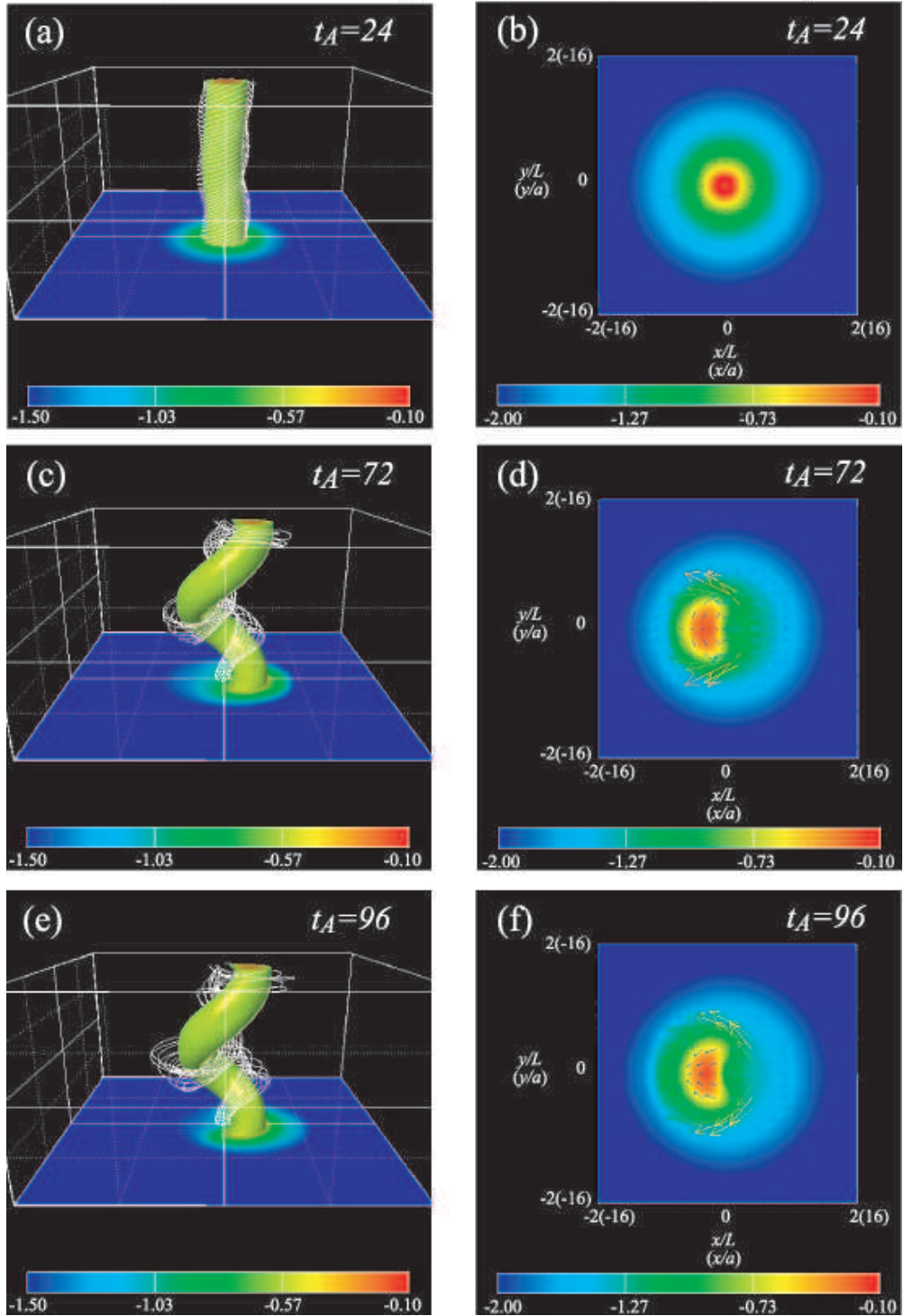}
\caption{3D density isosurface with transverse
slices at $z=0$ (left) and 2D transverse slices at the grid midplane
$z=L_{z}/2$ (right) for case A1n. Colors show the logarithm of the
density with magnetic field (white) lines. \label{f4}}
\end{figure}

\begin{figure}[hp]
\epsscale{0.8}
\plotone{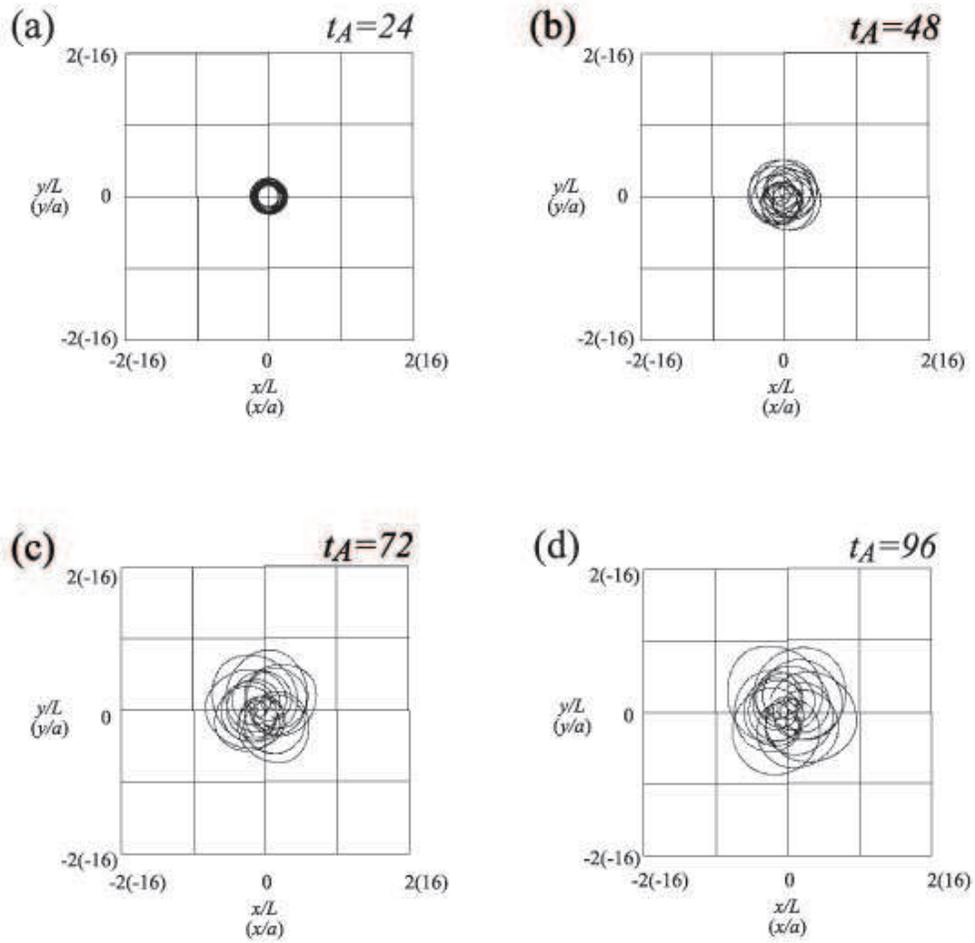}
\caption{\baselineskip 12pt Magnetic field structure seen from a pole
on view for case A1n. \label{f5}}
\end{figure}

\begin{figure}[hp]
\epsscale{0.8}
\plotone{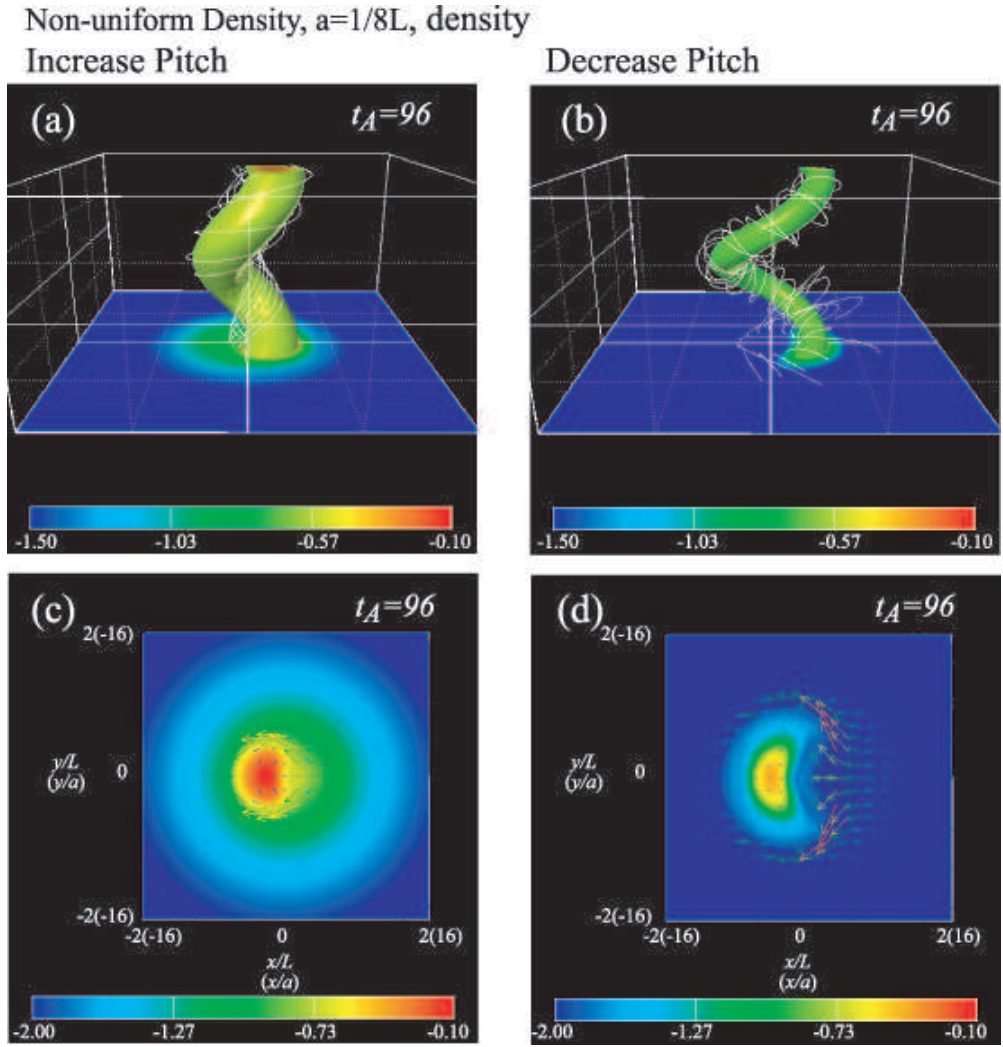}
\caption{\baselineskip 12pt 3D density isosurface with transverse slices at $z=0$ (upper panels) and 2D transverse slices at the grid midplane $z=L_{z}/2$ (lower panels) for case A2n (left) and A3n (right). Colors show the logarithm of the
density with magnetic field (white) lines.  \label{f6}}
\end{figure}

\begin{figure}[hp]
\epsscale{0.8}
\plotone{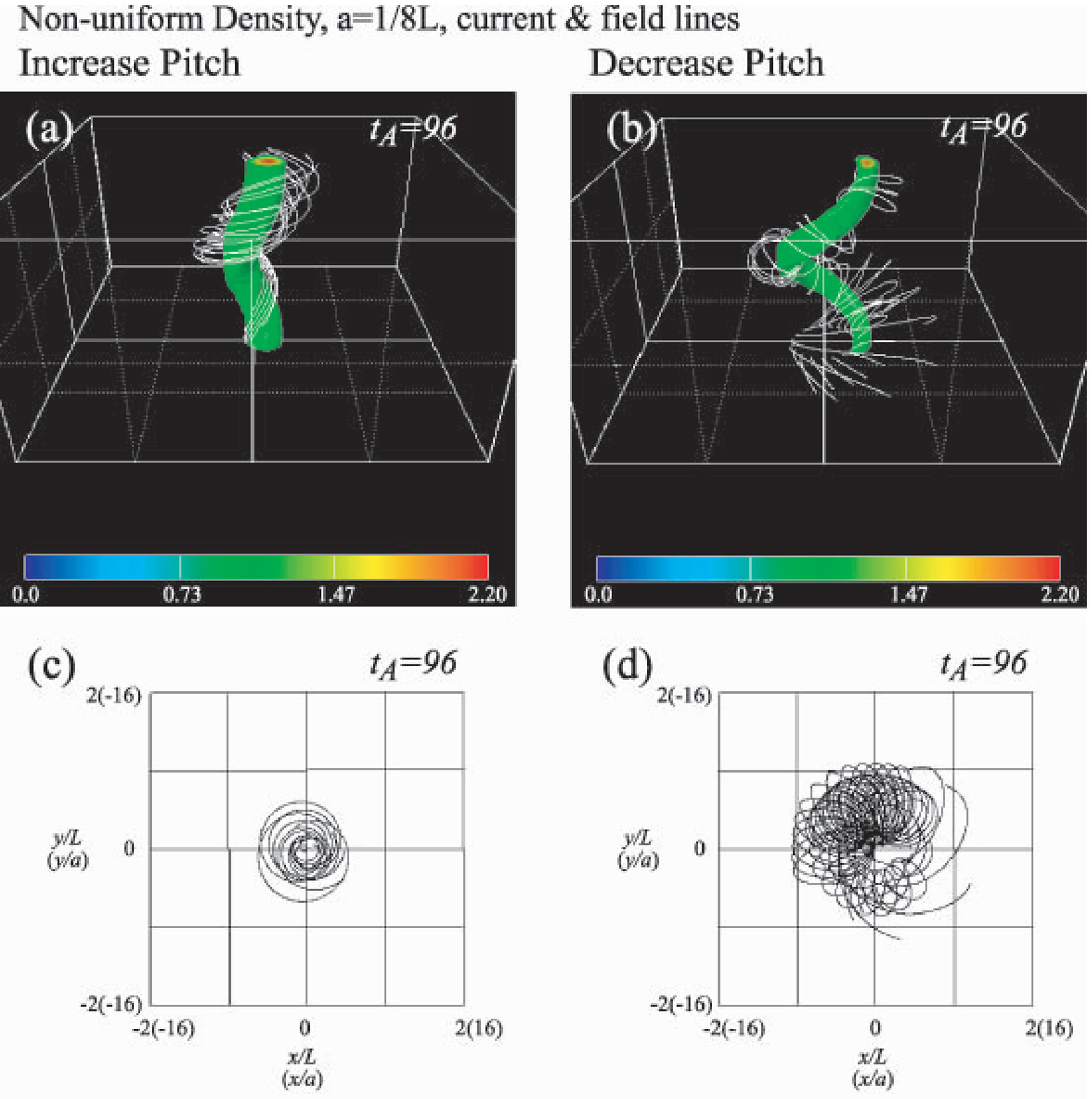}
\caption{3D axial current, $j_{z}$, isosurface (upper
panels) with magnetic field (white) lines and (bottom panels) magnetic
field structure seen from a pole on view for cases A2n (left) and A3n
(right). \label{f7}}
\end{figure}

Figure 6 shows density isosurfaces and transverse density slices at
the grid midplane, $z=L_{z}/2$, for the cases A2n
(increasing pitch) and A3n (decreasing pitch) that can be compared to
Figure 4 for case A1n with constant pitch.  The 3D
density structure from A2n looks similar to that from A1n.
The CD kink instability grows exponentially initially.  The
transverse density slice at the grid midplane (Fig.\ 6c) shows little
outwards motion of the high density region in the non-linear stage
similar to the constant pitch case A1n, however, there is little
outwards motion in the low density sheath surrounding the high density
region. This result is somewhat different from the constant pitch case
A1n and suggests a significant reduction in kink amplitude growth.
Results from the decreasing pitch case A3n are very
different. Figure 6b shows a more slender helical density structure
wrapped by the magnetic field. Recall that transition from linear to
non-linear evolution was reached at a much earlier time, $t_S
\sim 50 t_{An}$ (see Fig.\ 3), than is shown here.  While the density cross
section (Fig.\ 6d) is similar to that of the constant pitch case A1n
(Fig.\ 4f), radial motion continues in the non-linear stage to the end
of the simulation.

Figure 7 shows the axial current isosurface and the magnetic field
line displacement seen from the pole-on view for cases A2n and A3n.
Like the constant pitch decreasing density case A1n, the axial current
twist is cospatial with the high density twist. In the non-linear
stage, the helically twisted axial current does not change
significantly in case A2n. A pole on view of field lines anchored
initially at $R/L=0.2$ at the $z=0$ surface (see Fig.\ 7c) shows that
the magnetic field displacement does not increase significantly in the
non-linear stage. This along with the density results shown in Figure
6a \& c suggest significant non-linear stabilization of the kink mode
in the presence of increasing pitch.  On the other hand, the axial
current isosurface for the decreasing pitch case A3n seen in Figure 7b
makes a slender helical structure coincident with the high density
structure. Large displacement in the magnetic field can be seen in the
pole on view (Fig.\ 7d). 

\section{Summary and Discussion}

We have investigated development of the CD kink instability of
force-free helical magnetic equilibria in order to study the influence
of the initial configuration on the linear and nonlinear evolution. We
followed the time evolution where the plasma column was perturbed
to excite the $m=\pm 1$ azimuthal mode. The kink developed
initially as predicted by linear stability analysis. The rate at which
the kink developed and, in particular, the non-linear behavior depended
on the density and magnetic pitch radial profile. For a constant
magnetic pitch profile a density decline with radius led to faster
linear growth and made a transition to the non-linear stage sooner
than occured for a uniform density as a result of the more gradual
decrease in the Alfv\'{e}n velocity with radius. While there were some
differences in the kink density structure in these two different
cases, amplitude growth of the kink continued to the terminal
simulation time in both cases.

More profound differences in development accompanied different magnetic
pitch profiles.  An increase in the magnetic pitch with radius led
to a reduced growth rate in the linear (exponential) growth stage and
made a transition to non-linear behavior at a later time when
compared to the cases with constant magnetic pitch.  In the nonlinear
stage, amplitude growth of the kink appeared to have ceased by the
terminal simulation time. On the other hand, a decrease in the
magnetic pitch led to more rapid growth in the the linear growth
stage and made a transition to non-linear behavior at an earlier time
when compared to the cases with constant magnetic pitch. In the
non-linear stage, a slender helical density and current carrying plasma
column wrapped by magnetic field developed and amplitude growth
continued to increase throughout the simulation.  Thus, the linear
growth and nonlinear evolution of the CD kink instability depended on
the radial density profile and strongly depended on the magnetic pitch
profile. Increasing magnetic pitch with radius was clearly stabilizing.

In all cases, the initial axisymmetric structure is strongly distorted
by the kink instability, even though not disrupted. It is important to
note that the instability develops relatively slowly, that is, the
characteristic time for the instability to affect strongly the initial
structure is roughly $\tau \sim 100~v_A/a$. The growth rate of the
kink instability is roughly $\kappa\sim 0.1~v_A/c$, the exact
coefficient being dependent on the structure of the undisturbed
state. Therefore it is not accidental that the full development of the
instability takes a long time.  In a jet context, our simulations
correspond to a perturbation that remains at rest in the flow
frame. Therefore, in order to check whether the instability would
affect a jet flow, one has to compare $\tau$ with the propagation
time.  In the relativistically moving case, time dilation slows the
instability further by the jet Lorentz factor so that the condition
for the instability to affect the jet structure may be written as
\begin{equation}
\frac z{c\gamma}> 100\frac a{v_A}.
\end{equation}
In relativistic jets, one could take $v_A \approx c$.
In order to find the final criterion, one has to know how the jet
radius, $a$, and Lorentz factor, $\gamma$, vary with the distance,
$z$. If the jet is narrow enough so that $\Omega a^2/c<z$, where
$\Omega$ is the angular velocity at the base of the jet, one can use
the scaling (Tschekhovskoy et al.\ 2008; Komissarov et al.\ 2009;
Lyubarsky 2009) $\gamma \sim \Omega a$. In this case one finds that
the criterion for the kink instability is
\begin{equation}
zc>100\Omega a^2.
\end{equation}
This means that the instability could affect the jet structure only if
the jet expands slowly enough. Assuming the parabolic shape for the
jet, $\Omega a/c=\xi (\Omega z/c)^k$, where
$k<1$ and $\xi\sim 1$ are dimensionless numbers, one finds that the
instability develops only if $k<1/2$. Even in this case, 
the characteristic scale
for the development of the instability is very large,
\begin{equation}
\Omega z/c\sim (10\xi)^{2/(1-2k)}.
\label{scale}
\end{equation}

The above estimates should be considered as preliminary because we
assumed that the jet moves as a whole so that there is a frame of
reference in which the plasma is at rest.  Clearly, the criterion
for the CD kink instability can be modified by the effects of
relativistic rotation, gradual shear, a surrounding sheath, and
sideways expansion. For example, in relativistic Poynting dominated
jets the growth rate of the instability strongly depends on the
transverse profile of the poloidal magnetic field so that $\tau$ goes
to infinity when the poloidal field is uniform (Istomin
\& Pariev 1994, 1996; Lyubarskii 1999). On the other hand, Poynting
dominated jets are accelerated such that the acceleration zone spans a
large range of scales (Tschekhovskoy et al.\ 2008; Komissarov et al.\
2008; Lyubarsky 2009). Therefore, a small kink growth rate does not
necessarily preclude significant growth of helical perturbations.  We
plan to perform 3D RMHD simulations including some of the missing physical
effects such as jet flow and rotation in future work. In particular,
this will allow us to investigate the interaction between CD and KH
driven structure in the relativistic regime.

\acknowledgments
Y.M., K.I.N., and P.H. acknowledge partial support by NSF AST-0506719,
AST-0506666, NASA NNG05GK73G, NNX07AJ88G, and NNX08AG83G. This work
is supported in part by US-Israeli Binational Science Foundation grant
2006170. The simulations were performed on the Columbia Supercomputer
at NAS Division at NASA Ames Research Center and the Altix3700 BX2 at YITP
in Kyoto University.

\newpage

\appendix
\section{Numerical Tests}

In the simulations presented here numerical viscosity and resistivity
are associated with the finite grid resolution. It is generally
expected that numerical viscosity affects the quantitative growth or
damping of instabilities and numerical resistivity allows unexpected
magnetic reconnection even when we assume the ideal MHD
condition. Therefore we perform several numerical tests to check
whether saturation is adequately captured and investigate how the
results in the non-linear evolution stage depend on numerical
resolution.

\begin{figure}[h!]
\epsscale{0.7}
\plotone{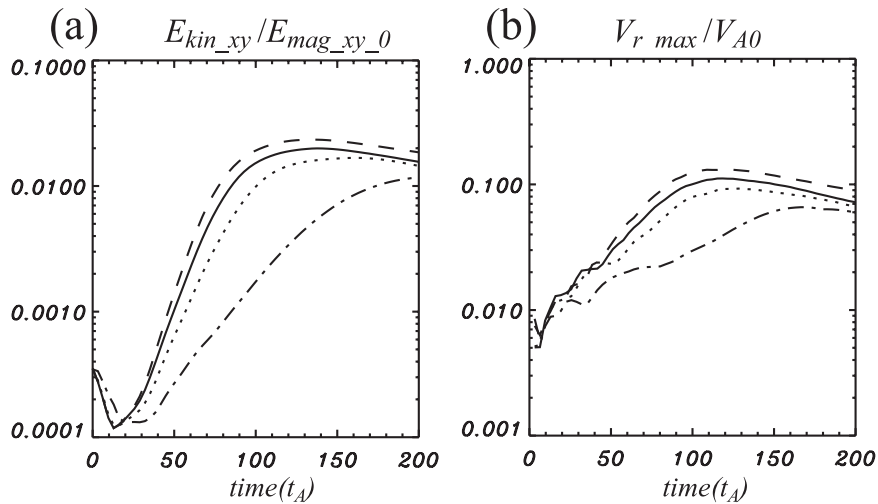}
\caption{Time evolution of (a) $E_{kin,xy}$
normalized by initial volume-averaged magnetic energy within $R/L \le
1.0$ and (b) maximum radial velocity normalzed by initial Alfv\'{e}n
velocity on the axis for case A1u with grid resolution of
$\Delta L = $ L/60 (dashed), L/40 (solid), L/30 (dotted), and L/20
(dash-dotted).
\label{fA1}}
\end{figure}
First, we repeat case A1u (uniform density with constant pitch) for
four grid resolutions from 20 to 60 computational zones per simulation
length unit $L= 8a$. Figure 8 shows the time evolution of
$E_{kin,xy}$ and the maximum radial velocity for different grid
resolutions. The results depend on grid resolution. 
Higher grid resolution shows faster growth and an earlier transition
time to the non-linear stage at a larger amplitude.  This result is
the same as that found in the resolution study of Baty \& Keppens
(2002). The lowest grid resolution with $\Delta L = L/20$ is clearly
inadequate. Although growth does depend on grid resolution, the
difference between our choice of $\Delta L = L/40$ and the highest
grid resolution of $\Delta L = L/60$ is not significant.  Therefore
our simulation results adequately capture the CD kink instability.

\begin{figure}[h]
\epsscale{0.7}
\plotone{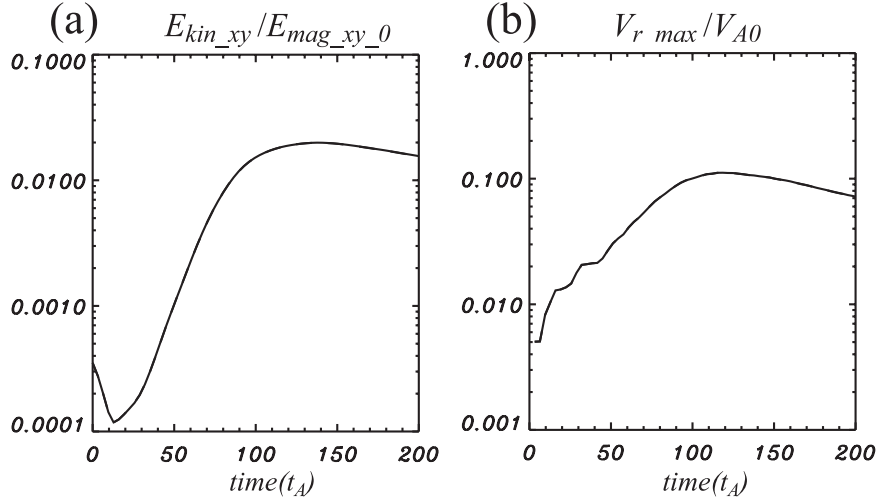}
\caption{Time evolution of (a) $E_{kin,xy}
$ normalized by initial volume-averaged magnetic energy within $R/L \le
1.0$ and (b) maximum radial velocity normalzed by initial Alfv\'{e}n
velocity on the axis for case A1u with transverse boundaries at
$x=y=\pm 2L$ ($\pm 16a$: solid) and $x=y= \pm 3L$ ($\pm 24a$:
dashed). \label{fA2}}
\end{figure}
Second, we check the influence of the transverse boundary. Figure 9
shows the time evolution of $E_{kin,xy}$ and the maximum radial
velocity within a box of length $L_{z}/2$ for a simulation box with
transverse boundaries at $x=y=\pm 2L$ ($\pm 16a$) and a simulation box
with transverse boundaries at $x=y=\pm 3L$ ($\pm 24a$). The results
for the time evolution of $E_{kin,xy}$ and the maximum radial velocity
are the same in both cases. We conclude that behavior of the CD kink
instability is not affected by our choice for the transverse boundary.

\begin{figure}[h!]
\epsscale{0.7}
\plotone{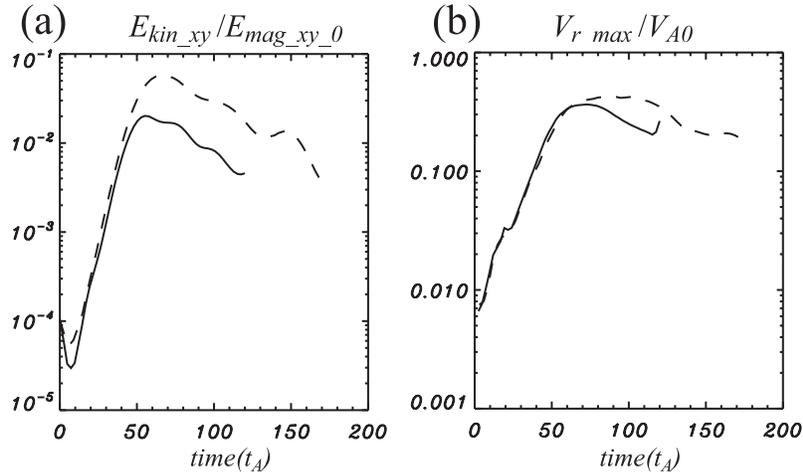}
\caption{Time evolution of (a) $E_{kin,xy} $
normalized by the initial volume-averaged magnetic energy within $R/L
\le 1.0$ and (b) the maximum radial velocity normalzed by the initial
axial Alfv\'{e}n velocity on the axis for case
A1n with an initial pressure of $p_{0}= 0.02 \rho_{0} c^{2}$ (solid)
and with a lower initial pressure of $p_{0}=0.005 \rho_{0} c^{2}$
(dashed). \label{fA3}}
\end{figure}
Third, we check the influence of the initial gas pressure. In the
force-free model we ignore the inertia and pressure of the plasma.
However, in the simulation the relativistic ideal MHD equations
are solved numerically not the force-free equations.  
Figure 10 shows the time evolution of $E_{kin,xy}$ and the maximum
radial velocity in a box of length $L_{z}/2$ for an initial pressure
of $p_{0}= 0.02 \rho_{0} c^{2}$ and $p_{0} = 0.005 \rho_{0}
c^{2}$. The results of the time evolution of $E_{kin,xy}$ and the
maximum radial velocity show the same linear growth rate in both
cases. However, the lower initial pressure case reaches saturation at
a slightly later time with higher amplitude but with a larger decline
in the non-linear stage. Thus, our required non-zero initial gas
pressure does affect the saturation level and non-linear development
of the CD kink instability somewhat.


\begin{thebibliography}{}

\bibitem[Appl et al.(2000)]{App00} Appl, S., Lery, T., \& Baty,
H. 2000, \aap, 355, 818

\bibitem[Barkov \& Komissarov(2008)]{Bar08} Barkov, M.V. \&
Komissarov, S.S. 2008, \mnras, 385, L28

\bibitem[Bateman(1978)]{Bat78} Bateman, G. 1978, MHD instabilites
(Cambridge, Mass., MIT Press, 1978, p.270)

\bibitem[Baty(2005)]{Bat05} Baty, H. 2005, \aap, 430, 9

\bibitem[Baty \& Keppens(2003)]{BK03} Baty, H. \& Keppens, R. 2002,
\apj, 580, 800

\bibitem[Beckwith et al.(2008)]{Bec08} Beckwith, K., Hawley, J.F., \&
Krolik, J.H. 2008, \apj, 678, 1180

\bibitem[Begelman(1998)]{Beg98} Begelman, M.C. 1998, \apj, 493, 291

\bibitem[Begelman et al.(1980)]{Beg80} Begelman, M. C., Blandford,
R.D., \& Rees, M.J. 1980, \nat, 287, 307

\bibitem[Blandford(1976)]{Bla76} Blandford, R.D. 1976, \mnras, 176, 465

\bibitem[Blandford \& Payne(1982)]{Bla82} Blandford, R.D. \& Payne, D.G.
1982, \mnras, 199, 883

\bibitem[Blandford \& Znajek(1977)]{Bla77} Blandford, R D. \& Znajek, R.L.
1977, \mnras, 179, 433

\bibitem[Bromberg \& Levinson(2008)]{Bro08} Bromberg, O. \& Levinson,
A. 2008, preprint (arXiv:0810.0562)

\bibitem[Carey \& Sovinec(2009)]{Car09} Carey, C.S. \& Sovinec,
C.R. 2009, preprint (arXiv:0902.0369)

\bibitem[De Villiers et al.(2003)]{DeV03} De Villiers, J.-P., Hawley, J.F.,
\& Krolik, J.H. 2003, \apj, 599, 1238

\bibitem[De Villiers et al.(2005)]{DeV05} De Villiers, J.-P., Hawley, J.F.,
Krolik, J.H., \& Hirose, S. 2005, \apj, 620, 878

\bibitem[Ferrari(1998)]{Fer98} Ferrari, A. 1998, \araa, 36, 539

\bibitem[Fukue(1990)]{Fuk90} Fukue, J. 1990, \pasj, 42, 793

\bibitem[Georganopoulos \& Kazanas(2003)]{Geo03} Georganopoulos, M. \&
Kazanas, D. 2003, \apj, 594, L27

\bibitem[Ghisellini et al.(2005)]{Ghi05} Ghisellini, G., Tavecchio,
F., \& Chiaberge, M. 2005, \aap, 432, 401

\bibitem[Giannios et al.(2009)]{Gia09} Giannios, D., Uzdensky, D.A.,
\& Begelman, M.C. 2009, \mnras, accepted (arXiv:0901.1877)

\bibitem[G\'{o}mez et al.(2001)]{Gom01} G\'{o}mez, J.-L., Marscher,
A.P., Antonio, A., Jorstad, S.G.,\& Agudo, I. 2001, \apj, 561, L161

\bibitem[Hardee(2004)]{Har04} Hardee, P.E. 2004, \apss, 293, 117

\bibitem[Hardee(2007)]{Har07} Hardee, P.E. 2007, \apj, 664, 26

\bibitem[Hardee et al.(2007)]{Har07a} Hardee, P., Mizuno, Y., \&
Nishikawa, K.-I. 2007, \apss, 311, 283

\bibitem[Hawley \& Krolik(2006)]{HK06} Hawley, J.F. \& Krolik,
J.H. 2006, \apj, 641, 103

\bibitem[Istomin \& Pariev(1994)]{IP94} Istomin, Y. N. \& Pariev,
V.I. 1994, \mnras, 267, 629

\bibitem[Istomin \& Pariev(1996)]{IP96} Istomin, Y.N. \& Pariev,
V.I. 1996, \mnras, 281, 1

\bibitem[Koide(2003)]{Koi03} Koide, S. 2003, \prd, 67, 104010

\bibitem[Komissarov(2005)]{Kom05} Komissarov, S.S. 2005, \mnras, 359, 801

\bibitem[Komissarov \& Barkov(2009)]{Kom09} Komissarov, S.S., \&
Barkov, M. V. 2009, \mnras, submitted (arXiv:0902.2881)

\bibitem[Komissarov \& McKinney(2007)]{Kom07a} Komissarov, S.S., \&
McKinney, L.C. 2007, \mnras, 377, L49

\bibitem[Komissarov et al.(2007)]{Kom07b} Komissarov, S.S., Barkov,
M.V., Vlahakis, N., \& K\"{o}nigl, A. 2007, \mnras, 380, 51

\bibitem[Komissarov et al.(2008)]{Kom08} Komissarov, S.S., Vlahakis,
N., K\"{o}nigl, A., \& Barkov, M.V. 2008, \mnras, submitted
(arXiv:0811.1467)

\bibitem[K\"{o}nigl \& Choudhuri(1985)]{Kon85} K\"{o}nigl, A. \&
Choudhuri, A.R. 1985, \apj, 289, 173

\bibitem[Lery et al.(2000)]{LB00} Lery, T., Baty, H., \& Appl,
S. 2000, \aap, 355, 1201

\bibitem[Liu et al.(2007)]{Liu07} Liu, Y.T., Shapiro, S.L., \&
Stephens, B.C. 2007, \prd, 76, 084017

\bibitem[Lobanov \& Zensus(2001)]{Lob01} Lobanov, A.P. \& Zensus,
J.A. 2001, Science, 294, 128

\bibitem[Lovelace(1976)]{Lov76} Lovelace, R.V.E. 1976, \nat, 262, 649

\bibitem[Lyubarskii(1999)]{Lyu99} Lyubarskii, Y.E. 1999, \mnras, 308, 1006

\bibitem[Lyubarsky(2009)]{Lyu09} Lyubarsky, Y. 2009, \apj, submitted (arXiv:0902.3357)

\bibitem[McKinney(2005)]{McK05} McKinney, J.C. 2005, \apj, 630, L5

\bibitem[McKinney(2006)]{McK06} McKinney, J.C. 2006, \mnras, 368, 1561

\bibitem[McKinney \& Blandford(2008)]{McK08} McKinney, J.C., \&
Blandford, R.D., 2008, \mnras, submitted (arXiv:0812.1060)

\bibitem[McKinney \& Gammie(2004)]{McK04} McKinney, J.C., \& Gammie,
C.F. 2004, \apj, 977

\bibitem[McKinney \& Narayan(2007a)]{McK07a} McKinney, J.C., \&
Narayan, R. 2007a, \mnras, 375, 513

\bibitem[McKinney \& Narayan(2007b)]{McK07b} ----. 2007b, \mnras, 375, 531

\bibitem[Meier(2005)]{Mei05} Meier, D.L. 2005, \apss, 300, 55

\bibitem[Meier et al.(2001)]{Mei01} Meier, D.L. Koide, S., \& Uchida, Y. 
2001, Science, 291, 84

\bibitem[M\'{e}sz\'{a}ros(2006)]{Mes06} M\'{e}sz\'{a}ros, P. 2006,
Rep.Prog.Phys., 69, 2259

\bibitem[Mirabel \& Rodriguez(1999)]{mr99} Mirabel, I.F., \&
Rodriguez, L.F. 1999, \araa, 37, 4

\bibitem[Mizuno et al.(2004a)]{Miz04a} Mizuno, Y., Yamada, S., Koide,
S., \& Shibata, K. 2004a, \apj, 606, 395

\bibitem[Mizuno et al.(2004b)]{Miz04b} ----. 2004b, \apj, 615, 389

\bibitem[Mizuno et al.(2006)]{Miz06} Mizuno, Y., Nishikawa, K.-I.,
Koide, S., Hardee, P., \& Fishman, G. J. 2006, in Proc. VI Microquasar
Workshop: Microquasars and Beyond, ed. Tomaso (Trieste: SISSA), 45

\bibitem[Moll et al.(2008)]{Mol08} Moll, R., Spruit, H. C., \&
Obergaulinger, M. 2008, \aap, 492, 621

\bibitem[Nagataki(2009)]{Nag09} Nagataki, S. 2009, \apj, submitted (arXiv:0902.1908)

\bibitem[Nakamura \& Meier(2004)]{nm04} Nakamura, M., \& Meier,
D.L. 2004, \apj, 617, 123

\bibitem[Nakamura et al.(2007)]{Nak07} Nakamura, M., Li, H., \& Li,
S. 2007, \apj, 656, 721

\bibitem[Narayan et al.(2007)]{Nar07} Narayan, R., McKinney J.C., \&
Farmer, A.J. 2007, \mnras, 375, 548

\bibitem[Narayan et al.(2009)]{Nar09} Narayan, R., Li, J., \&
Tchekhovskoy, A. 2009, \apj, submitted (arXiv:0901.4775)

\bibitem[Ouyed et al.(2003)]{Ouy03} Ouyed, R., Clarke, D.A., \&
Pudritz, R.E. 2003, \apj, 582, 292

\bibitem[Penrose(1969)]{Pen69} Penrose, R. 1969, Nuovo Cimento, 1, 252

\bibitem[Piran(2005)]{Pir05} Piran, T. 2005, Rev. of Mod. Phys., 76, 1143

\bibitem[Stephens et al.(2008)]{Ste08} Stephens, B.C., Shapiro,
S.L., \& Liu, Y.T. 2008, \prd, 77, 044001

\bibitem[Stern \& Poutanen(2008)]{Ste08} Stern, B.E. \& Poutanen,
J. 2008, \mnras, 383, 1695

\bibitem[Tchekhovskoy et al.(2008)]{Tch08} Tchekhovskoy, A., McKinney,
J.C., \& Narayan, R. 2008, \mnras, 388, 551

\bibitem[Tchekhovskoy et al.(2009)]{Tch09} ----. 2009, \apj, submitted (arXiv:0901.4776)

\bibitem[Tomimatsu et al.(2001)]{Tom01} Tomimatsu, A., Matsuoka, T.,
\& Takahashi, M. 2001, \prd, 64, 123003

\bibitem[Urry \& Padovani(1995)]{Urr95} Urry, C M. \& Padovani,
P. 1995, \pasp, 107, 903

\bibitem[Zhang \& M\'{e}sz\'{a}ros(2004)]{Zha04} Zhang, B. \&
M\'{e}sz\'{a}ros, P.  2004, Int.J.Mod.Phys., A19, 2385

\end{thebibliography}
\end{document}